\documentstyle[preprint,aps,epsf]{revtex}
\input psfig
\begin{document}
\newcommand{\br}{{\bf r}}
\newcommand{\e}{\epsilon}
\draft
\title {Quantum dot dephasing by edge states}
\author {Y. Levinson}
\address {Department of
Condensed Matter Physics, The Weizmann Institute of Science,
Rehovot 76100, Israel}

\date {3 July, 1999}
\maketitle
\begin {abstract}
We calculate the dephasing rate of an electron state in a pinched
quantum dot, due to Coulomb interactions between the electron
in the dot and electrons in a nearby voltage biased
ballistic nanostructure.
The dephasing is caused by nonequilibrium
time fluctuations of the electron
density in the nanostructure, which create random electric fields
in the dot. As a result, the electron level in the dot
fluctuates in time, and the coherent part of the resonant
transmission through the dot is suppressed.

\end {abstract}
\section{I\lowercase{ntroduction}}

The dephasing of electron states in Quantum Dots (QD) was considered
mainly in connection with weak localization phenomena,
see experiments \cite{Cl95,Bir95-98} and theory \cite {Siv94,Bla96}.
A different type of phenomenon in which  dephasing is important
is interference phenomenon in an  Aharonov-Bohm ring 
\cite{Ste90}. If a pinched QD  is embedded in one of the arms
of such a ring the transmission through this arm is supported
by a resonant electron state in  the QD.
The dephasing of this state \cite{Lev97}
 suppresses  the interference in
the ring, and this  can be observed as a decrease of of the oscillating part
of the ring  conductance \cite{Buk98}.

The dephasing is due to
electron-phonon or electron-electron interactions
of the QD electrons with some ``environment'',  which
can either be in equilibrium or driven out of it
by external forces.
In the experiment \cite{Buk98} the dephasing was due to the
capacitive
interaction of the QD  with a voltage biased point contact (PC),
and the amount of dephasing was dependent on the bias.
In a situation like this one can separate the equilibrium
dephasing, which depends only on the temperature 
of the environment,
from an additional dephasing which is due to voltages applied
to the environment. The theory concerning this experiment was given in
\cite{Lev97,Ale97}.

In  recent experiments \cite{Spr}  the nanostructure (NS), that was
capacitively coupled to the QD, was a multiterminal
2DEG device in a quantizing magnetic field .
We present in this paper a generalization of the theory
given in \cite{Lev97} that  takes into account the specific
effects appearing due to the complicated geometry,
and the chirality of the states in the NS (see Fig.1).
A similar problem was addressed in \cite{But99} using
a different approach, based on lumped mesoscopic circuit
elements.
 Broadening of electron transitions in self-assembled
QD's due to Coulomb
interactions with the electrons in the wetting layer
was considered in \cite{Usk99}.

\section{M\lowercase{odel}}

We consider a QD with  a single level $\e_{0}$,
that is  described by the Hamiltonian
$H_{QD}=\e_{0} c^{+}c $,
where $c^{+}$ is an operator creating an  electron in the QD state. 
The NS is a multiterminal junction in the 2DEG
described by the Hamiltonian
\begin{eqnarray}
H_{NS}=\int d\br \Psi^{+} (\br)H(\br)\Psi (\br),
\qquad
H(\br)=
{1 \over 2m}\left(-i\nabla- {e\over c}{\bf A}(\br)\right)^2+U(\br),
\label{HNS}
\end{eqnarray}
where  $U(\br)$ is the potential confining the 2DEG, 
${\bf A}(\br)$ is the vector potential of the external magnetic field
$(e<0, \hbar =1)$,
and $\Psi (\br)$ is the electron field operator.
The capacitive interaction between the QD and the NS, 
assumed to be weak, is
\begin{eqnarray}
H_{int}=c^{+}c \int d\br W(\br)\rho(\br),
\label{Hint}
\end{eqnarray}
where $\rho(\br)=\Psi^{+}(\br)\Psi(\br)$ is the electron density
operator,  and $W(\br)$ is a Coulomb interaction kernel.

The perturbation Eq.(\ref{Hint}) has a ``dual'' meaning.
If one combines $H_{NS}+H_{int}$ one can see that $W(\br)$
is the change of the confining potential $U(\br)$ 
due to one electron
occupying the QD state (when $<c^{+}c>=1$),
while combining  $H_{QD}+H_{int}$ one can see that
$\int d\br W(\br)\rho(\br)$ is the change of the  energy $\epsilon_{0}$
due to the Coulomb interaction of the electron in the QD
with the electron density in the NS.

\section{D\lowercase{ephasing rate notion}}

At low temperatures the dephasing is due to  electron-electron
interactions \cite{Alt82}.
To calculate the dephasing rate we use the method developed
in \cite{Lev97}. 
The electron density in the NS fluctuates in time and creates
a fluctuating potential in  the QD
which brings about  fluctuations
of the energy level $\e_{0}$. These fluctuations are given by
\begin{eqnarray}
\delta\e_{0}(t)=\int d\br W(\br)\delta\rho(\br,t),
\label{de}
\end{eqnarray}
where $\delta\rho(\br,t)\equiv\rho(\br,t)-\langle\rho(\br)\rangle$.
As a result, the correlator
$\langle\delta\epsilon_0 (t)\delta\epsilon_0 (0) \rangle$
is defined by the density-density correlator
$\langle\delta\rho(\br,t)\delta\rho(\br',0)\rangle$,
while $\langle\delta\e_0 \rangle=0$.

Consider resonant transmission through the QD for
electron  energies $\e$
close to $\e_{0}$. When the QD level does  not fluctuate the transmission 
amplitude $t(\e)$ contains the Breit-Wigner factor
$-i/[(\epsilon-\epsilon_{0})+i\Gamma]$,
where $\Gamma$ is the width of the level due to 
the QD's connection with the leads.
 When the level fluctuates the transmission and reflection
can be elastic and inelastic. In  interference experiments
only the elastic transmission is of importance
 and to
obtain the  elastic transmission amplitude
$\langle t(\e)\rangle$
 one has to replace the Breit-Wigner factor by \cite{Lev97}
\begin{eqnarray}
\int_{0}^{\infty}dt\exp [-\Gamma t-\Phi(t)+i(\epsilon-\epsilon_{0})t],
\label{BWn}
\end{eqnarray}
where
\begin{eqnarray}
\Phi(t)={1\over 2}\int_{0}^{t}dt'\int_{0}^{t}dt''
\langle\delta\e_{0}(t')\delta\e_{0}(t'')\rangle.
\label{phi}
\end{eqnarray}
One can see that $|\langle t(\e)\rangle|<|t(\e)|$
which means dephasing of the QD state, 
that is responsible for resonant transmission. 
The same can be understood from the 
dynamics of the QD state amplitude \cite{Lev97},
$\langle c(t)^{+}c(0)\rangle =\langle c(0)^{+}c(0)\rangle
\exp[i\epsilon_{0}t-\Gamma t-\Phi(t)].$

The level fluctuations are characterized by their amplitude
$\langle(\delta \e_{0})^2\rangle^{1/2}$
and by the correlation time $\tau _{c}$.
The amplitude is proportional to the strength of the capacitive coupling
$W$,  while the correlation time is independent of $W$ 
and is determined by the correlation time of the
density-density correlator.
Hence,  for weak enough coupling,  one has
$\langle(\delta \e_{0})^2\rangle^{1/2}\tau _{c}\ll 1$,
which corresponds to dynamical narrowing. In this case
$\Phi(t)=\gamma t$ and  the integral Eq.(\ref{BWn})
reduces to
$-i/[(\epsilon-\epsilon_{0})+i(\Gamma+\gamma)]$.
Here
\begin{equation} 
\gamma=\pi K(0),
\label{gamma}
\end{equation}
with the following level oscillations' spectrum  
\begin{equation}
K(\omega)={1\over  2\pi}\int_{-\infty} ^{+\infty}
dt e^{i\omega t}
\langle\delta\epsilon_0 (t)\delta\epsilon_0 (0) \rangle.
\end{equation}
This result means that in the case of dynamical narrowing
one can describe the dephasing by a dephasing time 
$\tau_{\varphi}=\gamma ^{-1}$.
The dephasing rate can be estimated as 
$\gamma\simeq \langle(\delta \e_{0})^2\rangle\tau_{c}$,
and is smaller than the amplitude of the level fluctuations
$ \langle(\delta \e_{0})^2\rangle ^{1/2}$.
In the general case when 
$\langle(\delta \e_{0})^2\rangle^{1/2}\tau _{c}\agt 1$,
the transmission probability
$|\langle t(\e)\rangle|^2$
is not a Lorenzian,  and a dephasing time can not be defined.

\section{D\lowercase{ephasing rate calculation}}

To calculate the correlator $K(\omega)$ we
represent the field operator  in terms of scattering states (SS's) 
\cite{But92} (see Appendix)
\begin{eqnarray}
\Psi (\br)=\int{d\e \over 2\pi}
\sum_{\alpha n}a_{\alpha n}( \e)\chi_{\alpha n} (\e,\br),
\label{ss}
\end{eqnarray}
where $a_{\alpha n }^{+}(\e)$ is an operator creating an
incoming electron in channel $n$ of terminal $\alpha$,
with energy $\e$.
Performing calculations similar to those in \cite{Lev97}  we find
\begin{equation}
\label{Ks}
K(\omega)=\sum_{\alpha\alpha'}K^{\alpha\alpha'}(\omega),
\end{equation}
where the contribution from terminals $\alpha$ and $\alpha'$  is
\begin{equation}
K^{\alpha\alpha'}(\omega)=
{1\over 2}\int {d\e\over 2\pi}\int {d\e'\over 2\pi}
\sum_{nn'}f_{\alpha}(\e)[1-f_{\alpha'}(\e')]
|W_{\alpha n,\alpha' n'}(\e,\e')|^2
[\delta(\e'-\e-\omega)+\delta(\e'-\e+\omega)].
\label{Ko}
\end{equation}
Here we used the fact
that when the SS's are normalized to a unit of incoming
flux one has
$\langle a^{+}_{\alpha n }(\e)  a_{\alpha ' n'}(\e') \rangle
=2\pi\delta(\e -\e ')\delta_{\alpha n,\alpha ' n'}f_{\alpha}(\e)$,
where $f_{\alpha}(\e)$ is the Fermi distribution for energy $\e$
in terminal $\alpha$.
It is convenient to write it as $f(\e-\delta\mu_{\alpha})$,
where $\delta\mu_{\alpha}=\mu_{\alpha}-\mu$ and
$f(\e)=\left(e^{(\e-\mu)/T}+1\right)^{-1}$
is the Fermi distribution with some reference chemical potential $\mu$.
The matrix element entering Eq.(\ref{Ko}) contains SS's,
\begin{equation}
W_{\alpha n,\alpha' n'}(\e,\e')=\int d\br W(\br)
\chi_{\alpha n}(\e,\br)^*
\chi_{\alpha' n'}(\e',\br).
\label{ME}
\end{equation}
The integration here is over the interaction area, i.e.
over that part of the NS which is close enough to the QD
and where $W(\br)$ is not small (see Fig.1).

In what follows we consider the case
when the voltages $V_{\alpha}$ applied to all terminals
are small. We choose $\mu$ to be the equilibrium chemical potential
(when all $V_{\alpha}=0$) and $\delta\mu_{\alpha}=eV_{\alpha}$.
 In this case the relevant energies in Eq.(\ref{Ko}) 
correspond to the small energy window $|\e-\mu|\alt \max[T,eV]$,
where electron exchange between terminals happens.
We assume that within this energy window one can neglect the energy
dependence of the scattering states and hence of the matrix elements 
Eq.(\ref{ME}).
As a result we have
\begin{equation}
K^{\alpha\alpha'}(\omega)=
{1\over 2}|W_{\alpha ,\alpha' }|^2
\int {d\e\over 2\pi}\int {d\e'\over 2\pi}
f(\e-eV_{\alpha})[1-f(\e'-eV_{\alpha'})]
[\delta(\e'-\e-\omega)+\delta(\e'-\e+\omega)],
\label{Ko1}
\end{equation}
with an effective matrix element
\begin{equation}
|W_{\alpha ,\alpha'}|^2\equiv \sum_{nn'}|W_{\alpha n,\alpha' n'}|^2.
\label{EME}
\end{equation}

Shifting the integration variables in Eq.(\ref{Ko1})
by $eV_{\alpha}$ and $eV_{\alpha '}$
one can see that the diagonal contributions $K^{\alpha\alpha}$
do not depend on the applied voltages $V_{\alpha}$,
and are equal to their equilibrium values
at $V_{\alpha}=0$, i.e.

\begin{equation}
K^{\alpha\alpha}(\omega)=
{1\over 8\pi^2}|W_{\alpha ,\alpha }|^2 F_{T}(\omega),
\label{Kod}
\end{equation}
where
$F_{T}(\omega)=\omega\coth (\omega/ 2T)=
2\omega\left[N_{T}(\omega)+{1\over 2}\right]$, with
$N_{T}(\omega)=\left(e^{\omega/T}+1\right)^{-1}$.
Note that $F_{T}(0)=2T$ and $F_{0}(\omega)=|\omega|$.

For the nondiagonal contributions $\alpha\neq \alpha'$
we find after a  shift of the integration variables
\begin{equation}
\label{Kond}
K^{(\alpha\alpha')}(\omega)
\equiv K^{\alpha\alpha'}(\omega)+K^{\alpha'\alpha}(\omega)=
{1\over 8\pi^2}|W_{\alpha ,\alpha' }|^2
[F_{T}(\omega+eV_{\alpha\alpha'})+F_{T}(\omega-eV_{\alpha\alpha'})],
\end{equation}
where $V_{\alpha\alpha'}=V_{\alpha}-V_{\alpha'}$.

Using Eq.(\ref{Ks}) one can find the dephasing rate as
a sum over {\it single terminals} and {\it pairs of terminals},
\begin{equation}
\label{gammas}
\gamma=\sum_{\alpha}\gamma^{(\alpha)}+
\sum_{\alpha<\alpha'}\gamma^{(\alpha\alpha')}.
\label{gs}
\end{equation}
It follows from Eq.(\ref{Kod}) that a {\it single terminal}
contributes to the dephasing only if SS's
emitted from this terminal reach the interaction area,
and that this  is always an equilibrium contribution,
\begin{equation}
\label{gd}
\gamma^{(\alpha)}=\pi K^{\alpha \alpha}(0)=
{1\over 4\pi}|W_{\alpha ,\alpha }|^2 T.
\end{equation}
A  {\it pair of terminals} contribute to the dephasing only if SS's
emitted from both terminals overlap in the interaction area,
\begin{equation}
\label{gnd}
\gamma^{(\alpha\alpha')}=
K^{(\alpha\alpha')}(0)=
{1\over 4\pi}|W_{\alpha ,\alpha' }|^2 F_{T}(eV_{\alpha\alpha'}).
\end{equation}
When both terminals are at the same voltage,  the contribution
of this pair to dephasing  is an equilibrium one. 

Note that $F_{T}(\omega)$ contains the ``zero point fluctuations'',
but they do not contribute to the equilibrium  dephasing rate,
given by $K(\omega)$ at $\omega=0$, hence for
 zero temperature there is no equilibrium dephasing.

If one is interested in nonequilibrium dephasing
one has to look only at  pairs of terminals which are at different
voltages, and which send scattering states that overlap
in the interaction region. The nonequilibrium
contribution of such a pair is
\begin{equation}
\label{gV}
\gamma^{(\alpha\alpha')}_{V}
\equiv \gamma^{(\alpha\alpha')}-\gamma^{(\alpha\alpha')}|_{V=0}
={1\over 4\pi}
|W_{\alpha ,\alpha' }|^2[F_{T}(eV_{\alpha\alpha'})-F_{T}(0)].
\label{gV}
\end{equation}
For zero temperature it reduces to
\label{gV0}
\begin{equation}
\gamma^{(\alpha\alpha')}_{V}|_{T=0}={1\over 4\pi}
|W_{\alpha ,\alpha' }|^2|eV_{\alpha\alpha'}|.
\label{gV0}
\end{equation}

Using the dual property of the interaction between the QD and the NS,
we consider now $W(\br)$ as a small variation of the confining potential
$U(\br)$ due to an electron occupying the QD.
As a result the scattering matrix of the NS is changed
according to Eq.(\ref{dS}) from $S$ to $S+\delta S$.
 Using in addition  Eq.(\ref{SScc}) we  can express the matrix elements
Eq.(\ref{ME}) in terms of the scattering matrix variation
\begin{equation}
W_{\alpha n\e, \alpha' n'\e}=
i\sum_{\beta m}S_{\beta m,\alpha n}(\e)^*
\delta S_{\beta m,\alpha' n'}(\e).
\label{WS}
\end{equation}
Pinching the QD to the Coulomb blockade regime one can change
the number of electrons in the QD one by one and measure 
the variation of the conductance matrix $G_{\beta\alpha}$
due to an additional electron in the QD  \cite{Fie93}.
Since $G_{\beta\alpha}=
\sum_{mn}|S_{\beta m,\alpha n}|^2-\delta_{\beta,\alpha}$,
this is a way to measure (in case of simple enough NS geometry)
the variation $\delta S$ and the matrix elements $W$.
This procedure was performed experimentally  \cite{Buk98} 
for the  simplest  NS, being a one channel point contact.

\section{D\lowercase{ephasing versus current noise}}

Dephasing is closely related to current noise since 
current fluctuations are related to charge density
fluctuations by the continuity equation.
The results obtained in \cite{But92} for the current noise
can be presented in the following form
\begin{equation}
\langle\delta I_{\alpha}\delta I_{\alpha'}\rangle=
{e^2\over 8\pi ^2}
\sum_{\beta\beta '}A_{\beta\beta '}^{\alpha\alpha '}
F_{T}(eV_{\beta\beta '}).
\label{II}
\end{equation}
Here the left hand side is the $\omega=0$ Fourier component
of the current cross-correlator in terminals $\alpha$
and $\alpha'$, 
\begin{equation}
A_{\beta\beta '}^{\alpha\alpha '}=
\sum_{m m'}A^{\alpha}_{\beta m,\beta' m'}
A^{\alpha'}_{\beta' m',\beta m},
\label{}
\end{equation}
with
\begin{equation}
A^{\alpha}_{\beta m,\beta' m'}=
\delta_{\beta\alpha}\delta_{\beta'\alpha}\delta_{mm'}-
\sum_{n}S^*_{\alpha n,\beta m}S_{\alpha n,\beta' m'},
\label{}
\end{equation}
where the scattering matrix is at $\e=\mu$.
One can see that a  {\it single terminal} $\beta$
contributes  to 
$\langle\delta I_{\alpha}\delta I_{\alpha'}\rangle$
only if SS's emitted from this terminal
reach both terminals $\alpha$ and $\alpha'$,
and that this contribution, given by the term with $\beta'=\beta$
 is always equilibrium. 
 {\it Pairs of terminals} $\beta$ and $\beta'$
contribute to
$\langle\delta I_{\alpha}\delta I_{\alpha'}\rangle$
 only if SS's emitted from each of these terminals
reach both terminals $\alpha$ and $\alpha'$.
This contribution given by terms with $\beta'\neq\beta$
contains a nonequilibrium part.
These conditions are very similar to those in case of dephasing.

\section{E\lowercase{xamples and discussion}}

We consider first a simple one channel 2-terminal device with a 
gate and a QD
(see Fig.2). The scattering matrix of the gate (in the absence
of an electron in the QD) is 
\begin{equation}
\label{}
S=
\left|\begin{array}{cc}r&{\tilde t}\\t& {\tilde r}\end{array}\right|
\equiv
\left|\begin{array}{cc}\cos\theta e^{i\alpha}&\sin\theta e^{i{\tilde \beta}}
\\ \sin\theta e^{i \beta}&\cos\theta e^{i{\tilde \alpha}}
\end{array}\right|,
\qquad {\tilde\alpha}-{\tilde\beta}=\pi-(\alpha-\beta),
\end{equation}
where $r$ and $t$ correspond to reflection and transmission
of the SS's approaching the gate from left,
while ${\tilde r}$ and ${\tilde t}$ correspond to the SS's
approaching the gate from the right.
If the magnetic field $B\neq 0$ the scattering matrix is not
symmetric, $t\neq {\tilde t}$.

In case of zero temperature the equilibrium part of
the QD state dephasing rate vanishes, and the nonequilibrium part
is according to Eq.(\ref{gV0}) and Eq.(\ref{WS})
\begin{equation}
\label{}
\gamma=\pi|W|^2  eV,
\qquad\qquad
|W|^2=|r^*\delta {\tilde t}+t^*\delta{\tilde r}|^2=
|\delta\theta-i(\delta\beta- \delta \alpha )  
\sin\theta\cos\theta |^2.
\end{equation}
Here $V\equiv|V_{12}|$ and $\delta {\tilde t}$ , $\delta{\tilde r}$
are the changes of the transmission and the reflection amplitudes
due to the electron in the QD. 
This result was obtained in \cite{Lev97} for $B=0$ and a symmetric gate.
The shot noise in this device (in the absence of an electron in the QD) is
\cite{Les89} 
\begin{equation}
\label{}
\langle(\delta I)^2\rangle=(e^2/4\pi ^2)|r|^2|t|^2 eV.
\end{equation}

Both the dephasing and the shot-noise are due to the same
nonequilibrium fluctuations, but they are {\it not proportional}
to each other. 
To get some insight, consider first zero or a weak magnetic field,
when both SS's 1 and 2 occupy the whole crossection of the sample
(see Fig. 2).
If we assume  there is no reflection from  the gate,
i.e. $|t|^2=1$,  $|r|^2=0$, we find
 $\langle(\delta I)^2\rangle=0$,
while $\gamma=\pi |\delta r|^2 eV\neq 0$.
The dephasing is nonzero  because  SS's emitted from different
terminals overlap near the QD,
while the shot noise is zero because each terminal 
(where the shot noise is measured)
is feeded only by one SS.
The situation for the shot noise changes if the gate is reflecting,
in which case each terminal is feeded by both SS's.

One can understand this difference from the following
simple calculation.
In a channel without reflection the wave function is
$\psi(x)=ae^{ikx}+be^{-ikx}$, where the two terms
are SS's coming from the left and right terminals.
 The corresponding charge and
current densities are $\rho(x)=e\{|a|^2+|b|^2+(ab^*e^{i2kx}+c.c.)\}$
and $j(x)=ev_{k}\{|a|^2-|b|^2\}$. 
What is important  for nonequilibrium fluctuations is the overlap
of SS's coming from different terminals, i.e. terms
proportional to $ab$. Such terms do not exist in $j$ but do
exist in $\rho$. This is why the shot-noise is zero, 
while the dephasing rate is not.
The term $(ab^*e^{i2kx}+c.c.)$ is of quantum
origin. It means that in a quasiclassical situation,
when the number of channels is large, this term will
average out due to ``integration'' over $k$.
When the gate is reflecting, $|t|^2\neq 1$,  $|r|^2\neq 0$,
both $\rho(x)$ and $j(x)$ contain terms proportional
to $ab$. One can easily check it using, for example,
the wave function to the left of the barrier 
$\psi(x)=a[e^{ikx}+re^{-ikx}]+b{\tilde t}e^{-ikx}$.
It is also important to notice that for $\omega =0$
the current and charge fluctuations are not coupled by the
continuity equation.

Consider now the same device in a strong  magnetic field 
when the  SS's are edge states (ES's) localized near the boundaries.
We assume also that the QD is far from the gate and the interaction
region does  not reach ES2. In this situation due
to the chirality of ES1 the QD can change only the phases
of $t$ and ${\tilde r}$. As a  result 
$|W|^2=(\delta\beta)^2|t|^2|r|^2$, i.e. the dephasing rate
is proportional to the shot-noise.
This is because {\it for chiral states} the current and charge
densities are proportional, $j=\rho v_{k}$.
(The connection between dephasing and the phase of $t$ was
mensioned in \cite{Sto98}).

As a second  example we consider a 4-terminal
 device similar to that used in experiment \cite{Spr},
with a geometry as  shown in Fig.3.
The source S ($\alpha=1$) and the drain D ($\alpha=2$)
are used to bias the device.
Two  floating terminals, 
one down-stream from S to D ($\alpha=3$)
and one up-stream from D to S ($\alpha=4$)
are ``dephasors'' according to \cite{But86}.
Gate A regulates the source-drain current,
while  gates  B and C  block the floating terminals.
The QD is located far from gate B.
We assume there is only one LL at the Fermi energy and that
the ES's  at opposite edges are well 
separated and do not overlap.
We will be interested only in nonequilibrium dephasing and 
consider zero temperature.

The SS's  emitted from the up-stream floating terminal 4
do not reach the interaction region and hence this terminal
does  not contribute to the dephasing of the QD.
(In what follows it is assumed that this terminal is blocked).
 Only scattering states
emitted from terminals 1,2, and 3 overlap in the interaction region,
and hence in accordance with  Eq.(\ref{gammas}) one has
$
\gamma=\gamma^{(12)}+\gamma^{(23)}+\gamma^{(31)}.
$
Since the QD is located far from point contact B,
all SS's  in the interaction region
have the form of the same ES $w^{-}_{2}(\br)\equiv w(\br)$
with different amplitudes, i.e.
$
\chi_1=e^{i\phi_1} t_A r_B w,\;
 \chi_2= e^{i\phi_2}{\tilde r}_A r_B w,\;
\chi_3=e^{i\phi_3} {\tilde t}_B w.
$
Here  $r_A$, $t_A$ and ${\tilde r}_A$, ${\tilde t}_A$ 
are the reflection and transmission amplitudes for ES's
approaching A from left and from right.
$ r_{B}$, $ t_{B}$ and 
${\tilde r}_{B}$, ${\tilde t}_{B}$ 
correspond to ES's  approaching B from
above and below.
The phase factors $e^{i\phi}$ depend on the position of the QD.
 The relevant matrix elements Eq.(\ref{EME}) are

\begin{equation}
|W_{12}|^2=|t_Ar_B|^2 |r_Ar_B|^2 |W|^2,\;
 |W_{13}|^2=|t_B|^2 |t_Ar_B|^2 |W|^2,\;
 |W_{23}|^2=|t_B|^2 |r_Ar_B|^2 |W|^2,
\end{equation}
where
\begin{equation}
W=\int d\br W(\br)|w(\br)|^2.
\end{equation}
Using these matrix elements we find
\begin{eqnarray}
\gamma^{(12)}=A|t_A|^2|r_A|^2|r_B|^4|V_{12}|,\;
\gamma^{(23)}=A|t_B|^2|r_B|^2|r_A|^2|V_{23}|,\;
\gamma^{(31)}=A|t_B|^2|r_B|^2|t_A|^2|V_{13}|,
\end{eqnarray}
with the  constant  $A=(e/4\pi)|W|^2$.

When  terminal 3 is open, i.e. $r_B=0$, 
SS1 and SS2  are absorbed in this terminal
and then  the interaction region is reached only by SS3.
There is no overlap in the interaction region
of SS's  emitted from different terminals and 
as a result all the contributions to nonequilibrium dephasing rate
vanish.
When  terminal 3 is blocked, i.e. $t_B=0$ we find
$\gamma=\gamma^{(12)}=A|t_A|^2|r_A|^2 |V_{12}|\equiv\gamma_{0}$.

Since terminal 3 is floating $V_3$ is given by the condition
that the current entering this terminal is zero,
which leads to $V_3=V_1 |t_A|^2+ V_2 |r_A|^2$.
Using this one finds
\begin{equation}
\gamma=\gamma_0\;|r_B|^2\;(2-|r_B|^2).
\end{equation}
One can see from this result that $\gamma <\gamma _{0}$,
i.e. the floating terminal
suppresses  the nonequilibrium dephasing rate of the QD state.
This result is in agreement with experiment \cite{Spr}.
We would like to stress that the suppression
 is not because of dephasing 
the SS's coming to the interaction region.
The absolute values of the matrix elements 
that enter the expression of the  dephasing rate 
according to Eq.(\ref{gV0}) do not depend on the phases
of the SS's overlapping in the interaction region.
If one would simply destroy their phases it would not affect
the dephasing rate $\gamma$.
The floating terminal suppresses  $\gamma$
because it absorbs the SS's moving towards the interaction
region from {\it different terminals}.
It is important to have in mind that a  theory based on the 
representation Eq.(\ref{ss}) assumes that terminals absorb
incoming waves as black bodies, which means that terminals
have infinite capacitance.

It is instructive to compare the dephasing rate with the shot noise.
When terminal 3 is blocked the shot noise is known to be
$
\langle(\delta I_{1})^2\rangle=
\langle(\delta I_{2})^2\rangle=(e^2/4\pi ^2)e|V_{12}||t_A|^2|r_A|^2.
$
Using Eq.(\ref{}) one can see that
opening  terminal 3 does  not change $\langle(\delta I_{1})^2\rangle$
but suppress $\langle(\delta I_{2})^2\rangle$ by exactly
the same factor $|r_{B}|^2(2-|r_{B}|^2)$
as the dephasing rate.

\section*{A\lowercase{cknowlegements}}

I am thankful to Y.Imry and B.Spivak for valuable discussions.
I also thank  M.Heiblum and  D.Shprinzak who informed me
about their recent unpublished experiments.
The work was 
supported  by  the Israel Academy of Sciences and Humanities
and Ministry of Science of Israel.

\appendix
\section*{}

In this Appendix we list some useful
properties of the Green function, the scattering states 
and the scattering matrix, valid also when the magnetic field $B\neq 0$.

For each terminal $\alpha$, and given energy $\e$  we define outgoing 
waves $w_{\alpha n}^{+}(\e,\br)$
 and incoming waves $w_{\alpha n}^{-}(\e,\br)$, where 
$n$ is the mode number (see Fig.1).
 In case of a strong magnetic field $w^{\pm}$
are ES's  and $n$ is the LL number.
 The waves $w^{\pm}$ are normalized to carry a  unit flux over the 
cross section of the terminal.
Choosing  the gauge  $A_{x}=-By, A_{y}=0$
for a given terminal, where 
$x$ and $y$ are the longitudinal and transverse coordinates
in this terminal, one can represent the waves as follows:
$w_{ n}^{+}(\e,\br)=\exp[ik_{n}^{+}(\e)x]\phi_{n}^{+}(\e,y),\;\;
w_{n}^{-}(\e,\br)=\exp[ik_{n}^{-}(\e)x]\phi_{n}^{-}(\e,y)$.
   
In what follows we use ``hat'' to indicate the magnetic field
inversion. It means for example that if $w^{+}_{n}$ is an 
outgoing  wave for the field $B$ (i.e. an outgoing  ES  for LL $n$)
then ${\hat w}^{+}_{n}$ is an outgoing wave for field $-B$
(i.e. an outgoing ES for the same LL near the opposite 
boundary). It is easy to check that
$
w_{\alpha n}^{\pm}(\e,\br)^*={\hat w}_{\alpha n}^{\mp}(\e,\br)
$
or equivalently $k_{n}^{\pm}(\e)=-{\hat k}_{n}^{\mp}(\e)$
and $\phi_{n}^{\pm}(\e,y)={\hat \phi}_{n}^{\mp}(\e,y)^*$.

Different functions $\phi(y)$ corresponding to the same wave vector 
$k$
are eigenfunctions of the same Hamiltonian and are orthogonal.
This is not the case when  two functions $\phi_{1}(y)$ and $\phi_{2}(y)$
correspond to the same energy $\e$, but to
 different wave vectors $k_{1}$ and $k_{2}$.
In this case the ``orthogonality'' relations are  \cite{Suk99}
\begin{equation}
\int dy \phi_{1}\phi_{2}
[(k_{1}-{e\over c}A_{x})+(k_{2}-{e\over c}A_{x})]=0,
\label{ort1}
\end{equation}
and
\begin{equation}
\int dy \phi_{1}{\hat \phi}_{2}
[(k_{1}-{e\over c}A_{x})-(k_{2}+{e\over c}A_{x})]=0.
\label{ort2}
\end{equation}

For a given energy $\e$ the incoming field in terminal $\alpha$ 
is a superposition of incoming waves
$
\sum_{n}a_{\alpha n}(\e)w_{\alpha n}^{-}(\e,\br),
$
while the outgoing field in terminal $\beta$ is a superposition 
of outgoing waves
$
\sum_{m}b_{\beta m}(\e)w_{\beta m}^{+}(\e,\br).
$
The scattering matrix connects the amplitudes of the incoming and 
outgoing waves
\begin{equation}
b_{\beta m}(\e)=\sum_{\alpha n}S_{\beta m,\alpha n}(\e)
a_{\alpha n}(\e).
\end{equation}
The scattering matrix is unitary due to flux conservation
\begin{equation}
\sum_{\beta m}S_{\beta m,\alpha n}(\e)^*S_{\beta m,\alpha' n'}(\e)
=\delta_{\alpha n,\alpha' n'},
\end{equation}
and due to time reversal
\begin{equation}
S_{\beta m,\alpha n}(\e)={\hat S}_{\alpha n,\beta m}(\e).
\label{}
\end{equation}

A scattering state  $\chi_{\alpha n}(\e,\br)$ 
is defined as a solution of the Schroedinger equation with
energy $\e$ excited by an
incoming wave $w_{\alpha n}^{-}(\e,\br)$.
Complex conjugate scattering states are solutions of the 
Schroedinger equation with inverted magnetic field.
Comparing the behaviour of $\chi^*$ and ${\hat \chi}$
at infinity one finds that
\begin{equation}
\label{SScc}
\chi_{\alpha n}(\e,\br)^*=
\sum_{\beta m}S_{\beta m,\alpha n}(\e)^*
\;{\hat \chi}_{\beta m}(\e,\br)
\end{equation}
and also
\begin{equation}
{\hat \chi}_{\alpha n}(\e,\br)=
\sum_{\beta m}{\hat S}_{\beta m,\alpha n}(\e)
\; \chi_{\beta m}(\e,\br)^*.
\end{equation}

The Green function is defined by the equation
\begin{equation}
[H(\br)-\epsilon]G_{\epsilon}(\br,\br')=-\delta(\br-\br'),
\end{equation}
with the Hamiltonian given by Eq.(\ref{HNS}).
The boundary conditions are $G_{\epsilon}(\br,\br')=0$
when $\br$ is at the boundary of the NS, and
 these correspond to outgoing waves when $\br$
approaches infinity in some of its terminals.
The Green theorem in case $B\neq 0$ is as follows \cite{Lev90}
\begin{equation}
\int (\br) (uHv-vH^*u)=-{i\over 2m}\oint dl\; {\bf n}
\left[u(-i\nabla-{e\over c}{\bf A})v-
 v(-i\nabla+{e\over c}{\bf A})u  \right],
\end{equation}
where ${\bf n}$ is the unit normal vector directed outside the NS,
$dl$ is an element of the boundary.

Using this theorem and Eq.(\ref{ort2})
one can prove the symmetry
$
G_{\epsilon}(\br',\br)={\hat G}_{\epsilon}(\br,\br').
$

When $\br$ approaches infinity in terminal $\beta$
\begin{equation}
\chi_{\alpha n}(\e,\br)|_{\br\rightarrow \infty\beta}=
\delta_{\alpha\beta}w_{\alpha n}^{-}(\e,\br)+
\sum_{m}S_{\beta m,\alpha n}(\e)w_{\beta m}^{+}(\e,\br),
\end{equation}
and
\begin{equation}
\label{Ginf}
G_{\epsilon}(\br,\br')|_{\br\rightarrow \infty\beta}=
-i\sum_{m}w_{\beta m\e}^{+}(\br){\hat \chi}_{\beta m\e}(\br').
\end{equation}
The first equation follows from  the definition of the
scattering states and the scattering matrix,  while the second
equation can be obtained as follows. From the explicit expression
of the Green function for a waveguide  in a magnetic field 
given in \cite{Lev91}
one can see that a unit source $-\delta(\br-\br')$
at $\br'\rightarrow\infty\beta $ excites an incoming field
$
-i\sum_{m} w^{-}_{\beta m}(\e,\br')^*  w^{-}_{\beta m}(\e,\br).
$
Since each wave $ w^{-}_{\beta m}(\e,\br)$ excites a state
$\chi_{\beta m}(\e,\br)$ we find that

\begin{equation}
G_{\e}(\br,\br')|_{\br'\rightarrow \infty\beta}=
-i\sum_{m}w^{-}_{\beta m}(\e,\br')^*
\chi_{\beta m}(\e,\br).
\end{equation}
Using the symmetry of $G$, and the relation between $w^{-}$
and $w^{+}$, we find the relation given above.

A useful function is defined as follows
\begin{equation}
G_{\e}(\br,\br')-{\hat G}_{\e}(\br,\br')^*\equiv
-ig_{\e}(\br,\br').
\end{equation}
This function can be presented in terms of the scattering states
\begin{equation}
g_{\e}(\br,\br')=
\sum_{\alpha n}\chi_{\alpha n\e}(\br)\chi_{\alpha n\e}(\br')^*.
\end{equation}
Obviously $g_{\e}(\br,\br)$ is the local density of states.
Inverting $B$ one finds
$
{\hat g}_{\e}(\br,\br')=g_{\e}(\br,\br')^*.
$
For $B=0$ the function $g_{\e}(\br,\br')$ is real.

Let the confining potential $U(\br)$ be  subjected to some variation
$\delta U(\br)$. The variation of the scattering states
$\delta\chi_{\alpha m}(\e,\br)$ contains  only
outgoing waves and can be found from the first Born approximation
using the retarded Green function corresponding to
 the potential $U(\br)$.
The asymptotic behaviour of $\delta\chi_{\alpha m}(\e,\br)$
at $\br\rightarrow\infty\beta$ can  then be  found using Eq.(\ref{Ginf}).
As a result the variation of the scattering matrix is

\begin{equation}
\label{dS}
{\delta\over\delta U(\br)}S_{\alpha n,\beta m}(\e)=
-i{\hat \chi}_{\alpha n}(\e,\br)\chi_{\beta m}(\e,\br).
\end{equation}

\begin{figure} 
\centerline{\psfig{figure=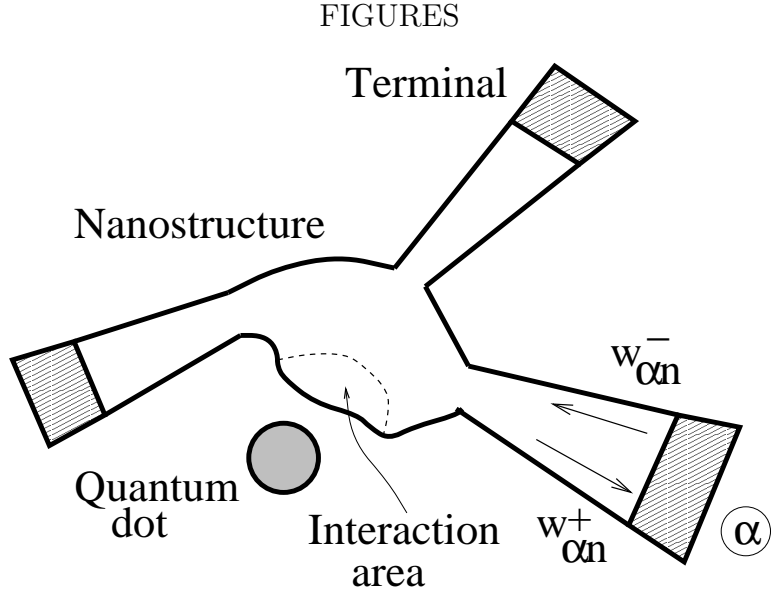,width=4in}}
\vspace{1cm} 
\caption{
 The quantum dot and the nanostructure. 
$w^{\pm}_{\alpha n}$ are waves
emitted and absorbed by terminal $\alpha$ (see text).}
\end{figure}

\begin{figure} 
\centerline{\psfig{figure=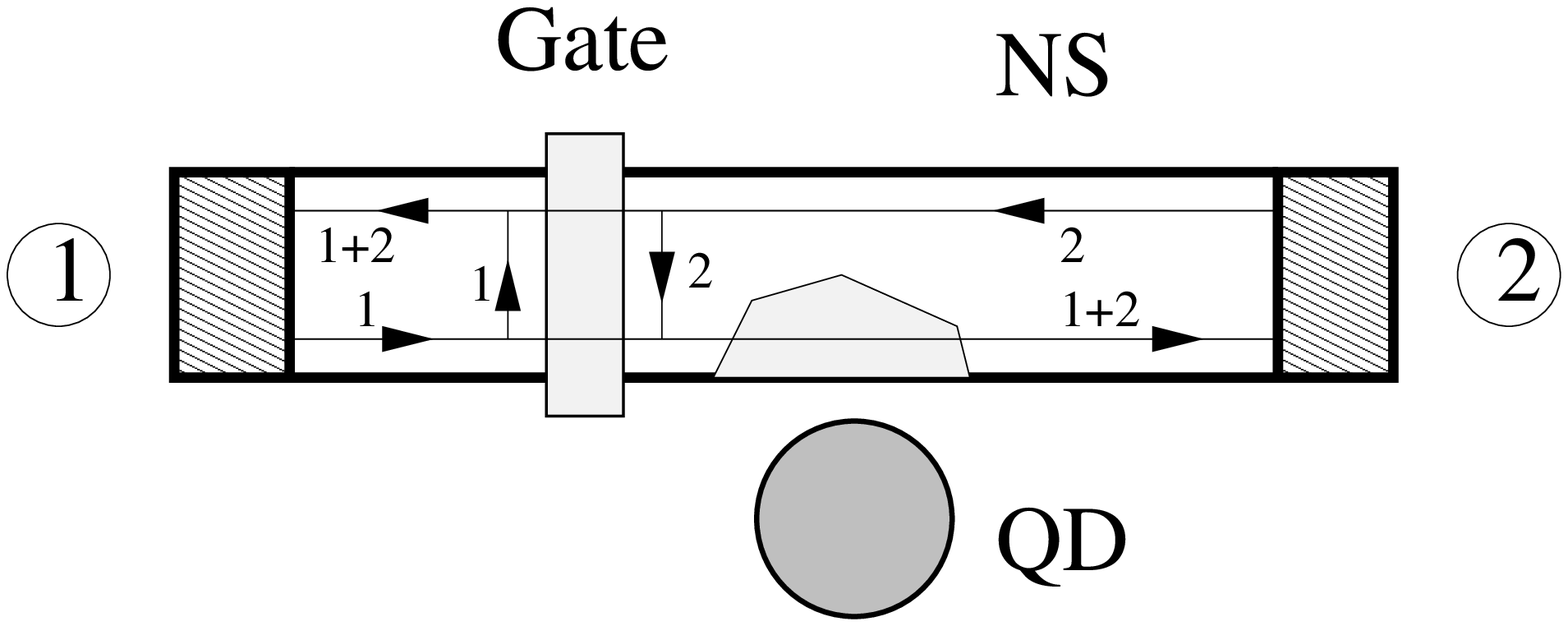,width=5in}}
\vspace{1cm}
\caption{ A two-terminal device. Scattering states
emitted from the terminals are labeled by the terminals'
corresponding numbers.
The interaction area is shown.}
\end{figure}

\begin{figure} 
\centerline{\psfig{figure=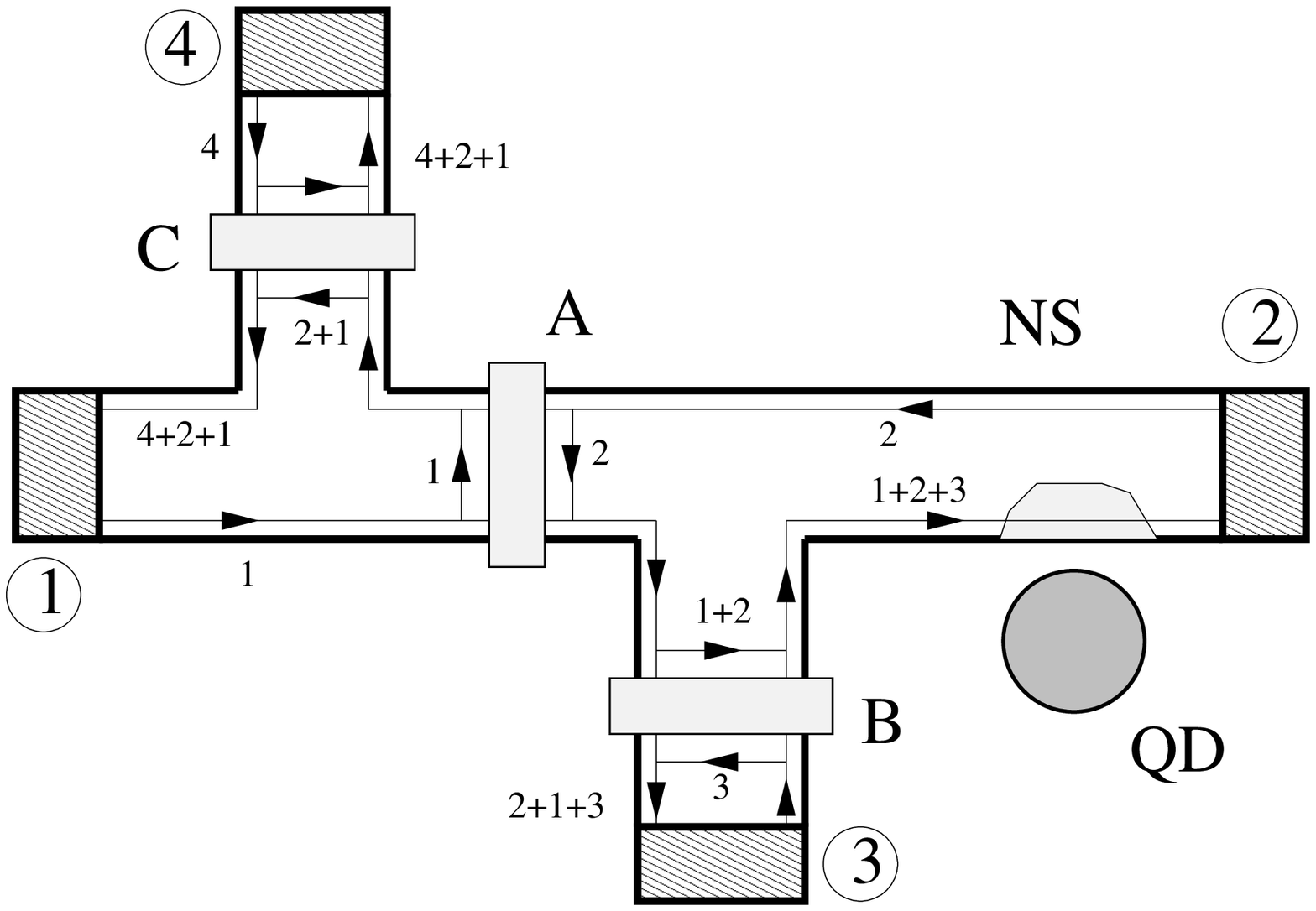,width=5in}}
\vspace{1cm}
\caption{ A four-terminal device.
Scattering states
emitted from the terminals are labeled by the terminals'
corresponding numbers.
The interaction area is shown.}
\end{figure}

\begin{references}
\bibitem{Cl95}
R.M.Clark {\it et al.},
Phys.Rev. B {\bf 52}, 2656, 1995

\bibitem{Bir95-98}
J.P.Bird {\it et al.},
Phys.Rev. B {\bf 51 }, 18036, 1995;
J.Phys.: Condens. Matter {\bf 10}, L55,1998 

\bibitem{Siv94}
U.Sivan, Y.Imry and A.Aronov,
Europhys.Lett., {\bf 28},115, 1994

\bibitem{Bla96}
Ya.M.Blanter,
cond-mat/9604101

\bibitem{Ste90}
A.Stern,Y.Aharonov and Y.Imry,
Phys.Rev. {\bf 41}, 3436, 1990

\bibitem{Lev97}
Y.Levinson,
Europhys.Lett., {\bf 39},299, 1997

\bibitem{Buk98}
E.Buks {\it et al.}
Physica {\bf 249-251}, 295, 1998
  
\bibitem{Ale97}
I.L.Aleiner, N. S.Wingreen and Y.Meir,
Phys.Rev.Lett. {\bf 79 }, 3740, 1997

\bibitem{Spr}
D.Sprinzak and M.Heiblum, unpublished

\bibitem{But99}
M.Buttiker and A.M.Martin,
cond-mat/9902320


\bibitem{Usk99}
A.V.Uskov, K.Nishi and R.Lang,
Appl.Phys.Lett. {\bf 74}, 3081, 1999


\bibitem{Alt82}
B.L.Al'tshuler, A.G.Aronov and D.E.Khmelnitskii,
J.Phys.C{\bf 15}, 7367, 1982

\bibitem{Fie93}
M.Field {\it et al.},
Phys.Rev.Lett. {\bf 70 }, 1311, 1993

\bibitem{But92}
M.Buttiker, 
Phys.Rev. B {\bf 46 }, 12485, 1992


\bibitem{Les89}
G.B.Lesovik, JETP Lett. {\bf 49}, 592, 1989



\bibitem{Sto98}
L.Stodolsky,
quant-ph/9805081

\bibitem{But86}
M.Buttiker, 
Phys.Rev. B {\bf 33 }, 3020, 1986

\bibitem{Suk99}
E.V.Sukhorukov, unpublished

\bibitem{Lev90}
Y.B.Levinson and E.V.Sukhorukov,
Phys.Lett. {\bf 149A}, 167, 1990

\bibitem{Lev91}
Y.B.Levinson and E.V.Sukhorukov,
J.Phys.: Condens.Matter {\bf 3}, 7291, 1991 



\end{references}
\end{document}